\documentclass[manuscript]{aa}
\usepackage[varg]{txfonts}

\usepackage{graphicx}
\usepackage{amssymb, amsmath} 
\usepackage{color}
\usepackage[usenames,dvipsnames,table]{xcolor}
\bibpunct{(}{)}{;}{a}{}{,} 

\def\mnras{MNRAS}

\def\nat{Nature}

\def \nh {N${\rm _H}$}

\def \ferg {\mbox{erg cm$^{-2}$ s$^{-1}$}}
\def \hcm {\hbox {\ifmmode $ atom cm$^{-2}\else atom cm$^{-2}$\fi}}

\def\inte{{\em INTEGRAL}} 
\def\xmm{{\em XMM-Newton}}

\def\chan{{\em Chandra}}

\def\ATCA{{\em ATCA}}
\def\j11{\mbox{\object{IGR~J11014-6103}}}
\def\lhn{Lighthouse nebula} 
\def\SNR{\object{SNR MSH~11-61A}}
\def\guitar{\object{PSR~B2224+65}}
\def\morla{\object{PSR~J0357+3205}}

\def \apj {ApJ}
\def \apjl {ApJL}
\def \apjs {ApJS}
\def \aap {A\&A}

\begin{document}

\title{The long helical jet of the \lhn, \j11} \titlerunning{The
  helical jet of the \lhn}

\author {L. Pavan\inst{\ref{inst1}} \and
  P. Bordas\inst{\ref{inst1},\ref{inst2}} \and
  G. P{\"u}hlhofer\inst{\ref{inst2}} \and
  M. D. Filipovi\'c\inst{\ref{inst3}} \and A. De Horta\inst{\ref{inst3}}
  \and A. O' Brien\inst{\ref{inst3}} \and M. Balbo\inst{\ref{inst1}}
  \and R. Walter\inst{\ref{inst1}} \and E. Bozzo\inst{\ref{inst1}}
  \and C. Ferrigno\inst{\ref{inst1}} \and
  E. Crawford\inst{\ref{inst3}} \and L. Stella\inst{\ref{inst4}} }

\institute{ISDC Data Center for Astrophysics, Universit\'e de
  Gen\`eve, chemin d'Ecogia, 16, 1290 Versoix,
  Switzerland\\
  \email{Lucia.Pavan@unige.ch}\label{inst1} \and Institut f{\"u}r
  Astronomie und Astrophysik, Universit{\"a}t T{\"u}bingen, Sand 1,
  D-72076 T{\"u}bingen, Germany\label{inst2} \and Computational
  Astrophysics, Imaging \& Simulation School of Computing \&
  Mathematics, University of Western Sydney, Australia\label{inst3}
  \and INAF - Osservatorio astronomico di Roma, via di Frascati 33,
  Monte Porzio Catone, Roma\label{inst4} }

\authorrunning{L.Pavan et al.}

\date{Received 2 September 2013/ Accepted 12 January 2014}

\abstract{
  Jets from rotation-powered pulsars so far have only been observed in
  systems moving subsonically through their ambient medium and/or
  embedded in their progenitor supernova remnant (SNR).  Supersonic
  runaway pulsars are also expected to produce jets, but they have not
  been confirmed so far.}
{We investigated the nature of the jet-like structure associated with
  the \inte\ source \j11 (the ``\lhn''). The source is a neutron star
  escaping its parent \SNR\ supersonically at a velocity exceeding
  1000 km s$^{-1}$.}
{We observed the \lhn\ and its jet-like X-ray structure through
  dedicated high spatial resolution observations in X-rays (with
  \chan) and in the radio band (with \ATCA).}
{Our results show that the feature is a true pulsar's jet. It extends
  highly collimated over $\gtrsim$11pc, displays a clear
  precession-like modulation, and propagates nearly perpendicular to
  the system direction of motion, implying that the neutron star's
  spin axis in \j11\ is almost perpendicular to the direction of the
  kick received during the supernova explosion.}
{Our findings suggest that jets are common to rotation-powered
  pulsars, and demonstrate that supernovae can impart high kick
  velocities to misaligned spinning neutron stars, possibly through
  distinct, exotic, core-collapse mechanisms.}


\maketitle

\defcitealias{pavan11}{Paper I} \defcitealias{tomsick:2012eu}{T12}


\section{Introduction}
Pulsar wind nebulae (PWNe) powered by pulsars that are still embedded
in their progenitor supernova remnant (SNR) are typically seen as
extended cocoons, in some cases accompanied by two collimated jets
\citep{Weisskopf-2000,gaensler2006,kargalstev-pavlov-2008,durant2013}.
As of yet, jets have not been clearly identified from runaway pulsars
that are traveling supersonically through the interstellar medium,
which show otherwise cometary-like features formed by the confinement
of their winds by a bow-shock \citep{gaensler2006}.
Deep statistical studies have proved that pulsar's spin axes (and
hence jets, when present) are generally aligned with their proper
motion \citep{Weisskopf-2000,johnston2007,durant2013}, a relation that
has been explained by numerous theoretical models of asymmetric
supernova mechanisms \citep{spruit1998,Lai2001,Janka-2012}.
Jets in runaway systems could therefore be either disrupted by the
system motion or aligned with (and hidden by) the bow-shocked PWN
\citep{kargalstev-pavlov-2008}, making their detection difficult.

In a few cases only, puzzling jet-like X-ray features extending over
parsec-scales have been observed in association with high velocity
pulsars, e.g., in the Guitar nebula \citep[powered by pulsar
\guitar;][]{Hui:2012fk} and in Morla \citep[powered by
\morla;][]{deluca2013}. In the latter case, the jet-like feature is
aligned with the pulsar proper motion, and may correspond to a trail
of thermal emission from the shocked ambient medium
\citep{marelli2013}.  In the Guitar nebula, in contrast, the
collimated X-ray outflow is inclined by $\sim 118 \degr$ with respect
to the backwards PWN, which is seen in the optical band as a bright
H-$\alpha$ nebula \citep{Cordes:1993fk}. Although observations
performed at different epochs confirmed the link between this extended
X-ray feature and the pulsar \citep{johnson-wang-10}, its nature is
still under discussion, and different scenarios, including a truly
ballistic jet \citep{johnson-wang-10, Hui:2012fk} or a flow of
high-energy electrons diffusing into the ambient magnetic field
\citep{bandiera2008}, have been proposed.\\
\indent The \inte\ source \j11, which we call the \lhn, located
11\arcmin\ southwest of the \SNR\ \citep[estimated distance $7\pm
1$~kpc,][]{reynoso2006}, is a complex system that, in X-rays, displays
a point source, an elongated cometary tail, and perpendicular to this,
a prominent jet-like feature (see \citealp{pavan11}, hereafter
Paper~I; and \citealp{tomsick:2012eu}, T12 in the following). While
the previous studies already suggested a PWN scenario for the cometary
tail in \j11, the jet-like feature remained as yet uninterpreted due
to the poor statistics of the available data.  To unveil its real
nature, we obtained high spatial resolution observations of the \lhn\
with the \chan\ X-ray observatory and in the radio band with the
Australia Telescope Compact Array (\ATCA).

\section{Observations and data analysis}
\label{sec:observations}

\subsection{Chandra X-ray observation}
\label{sec:chandra}

We observed the \lhn\ with \chan\ for 50~ks on October 11, 2012
(obs.ID~13787).  The data have been analyzed with the Chandra
Interactive Analysis of Observations package \citep[\texttt{CIAO}
version 4.5;][]{ciao} and reprocessed with standard tools using the
most recently available calibration, as recommended by the \chan\
team.  We applied standard filters to the event file, on grade,
status, and good time intervals.  The observation was not affected by
strong background flares, resulting in
a final net exposure of 49.4 ks.\\
We performed the observation with the Advanced CCD Imaging
Spectrometer (ACIS-I) detector pointing at source
\object{2XMM~J110145.0-610140} (``PSR'' hereafter, see
Fig.~\ref{ima:flat-fielded}), with a moderate offset of 0.8\arcmin\
with respect to the nominal aim point, to optimize the results for the
imaging of the PSR itself, the cometary PWN, and jet structures.  All
components of the \lhn\ and a large portion of the nearby \SNR\ are
included in the total ACIS field of view (FOV, see dashed box in
Fig.~\ref{ima:flat-fielded}).

\subsubsection{Imaging}
\label{sec:imaging}

\begin{figure*}[h!bt]
  \centering
  \includegraphics[width=0.9\textwidth]{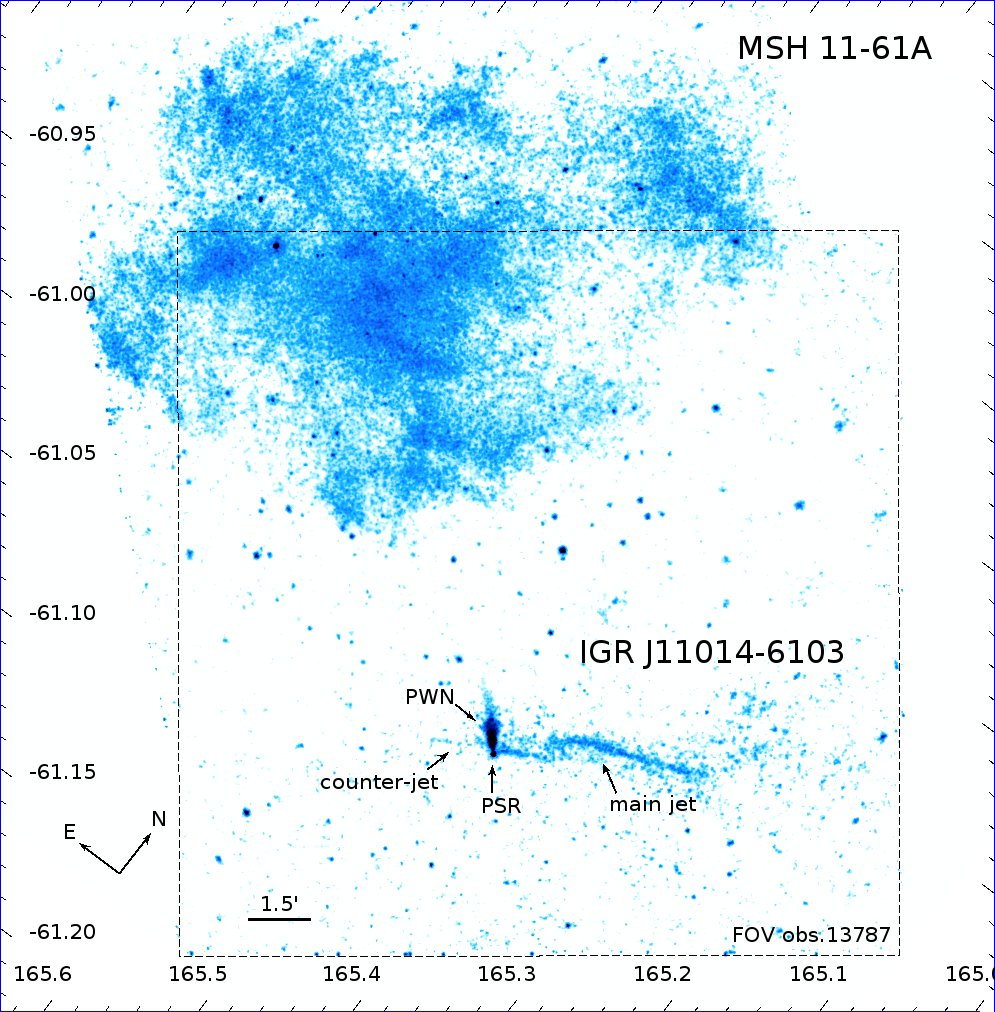}
  \caption{The \lhn\ (\j11) as viewed by \chan\ during our observation
    (obs.ID 13787, with the field of view indicated by the dashed
    box).  The image also includes archival observations of \SNR.  The
    two objects have likely been produced by the same supernova
    explosion 10~--~20 kyr ago.  The image is adaptively smoothed with
    a gaussian kernel, rotated by 37.5$\degr$, and normalized by the
    exposure to recover gaps between the chips.  Labels indicate the
    terms used throughout the text for the different components of the
    \lhn.}
  \label{ima:flat-fielded}
\end{figure*}

Chandra observations in imaging mode are performed with a dithering
technique that enables coverage of the gaps between CCDs, though these
are characterized by a reduced net exposure.
To recover these sections, we computed a fluxed image, by normalizing
the count map with the local net exposure.  The exposure map takes the
\chan\ effective area into account, and is therefore energy
dependent. We thus computed the instrument and exposure maps,
weighting them for the spectral distribution observed from the PSR
source (see Sect.~\ref{sec:spectra} and Table~\ref{tab:chanspec}),
through the standard \texttt{CIAO} tools
\texttt{make\_instmap\_weights} and \texttt{mkinstmap}. We verified
that different, reasonable weights for the instrument map did not
significantly affect the resulting corrected image
(Fig.~\ref{ima:flat-fielded}).
We detected X-ray extended emission up to 3~keV from the \SNR\ in
agreement with what has been previously reported by
\citet{Garcia:2012fk}.  The three components of the \lhn\ were clearly
visible in our \chan\ observation. We detected PSR at
\mbox{R.A. 11:01:44.915}, \mbox{DEC. -61:01:38.66} (J2000, associated
uncertainty 0.64\arcsec) in agreement with the position previously
reported by \citetalias{tomsick:2012eu}.  The PWN is detected up to
1.2\arcmin\ from PSR.  The main jet is clearly detected up to
$\sim$5.5\arcmin\ from PSR in a northwest direction, forming an angle
of $\sim$ 104$\degr$ with the PWN axis.  The \chan\ data also revealed
for the first time that the long jet is highly collimated and displays
a strikingly well-defined corkscrew modulation. In the corrected
image, the jet shows a remarkable change of orientation at about
1.4\arcmin\ from PSR. We did not attempt a more refined analysis and
interpretation of this region as it lies in a gap between two CCDs,
and thus is characterized by a much lower net exposure time with
respect to the surroundings.

In addition to these three components, our observation has revealed a
fainter linear region opposite the jet, in a southeast direction (see
Fig.~\ref{ima:flat-fielded}). This extended region is clearly detected
with the Voronoi Tessellation and Percolation source detection
(\texttt{CIAO/vtpdetect}) algorithm up to 1.5\arcmin\ from PSR, with a
detection significance of 3.7$\sigma$. The orientation and position of
this linear feature is readily interpreted as a counterjet.\\
The counterjet is well aligned with the direction of the main jet
within the first arcminute from PSR.  Both features intersect the PWN
at 7.4\arcsec\ from the point source.  Within 30\arcsec\ from the PSR,
however, the jet bends directly toward the point source and is
smoothly connected to it (see Fig.~\ref{ima:flat-fielded}).  The
alignment between the jet and counterjet, together with the change of
brightness at the PSR location and their smooth connection to PSR,
rule out any chance coincidence between these structures.

\subsubsection{Point-like source PSR}
\label{sec:psr}

\begin{figure}[h!tb]
   \centering
   \includegraphics[trim = 0mm 0mm 10mm 0mm, clip,
   width=0.3\textwidth]{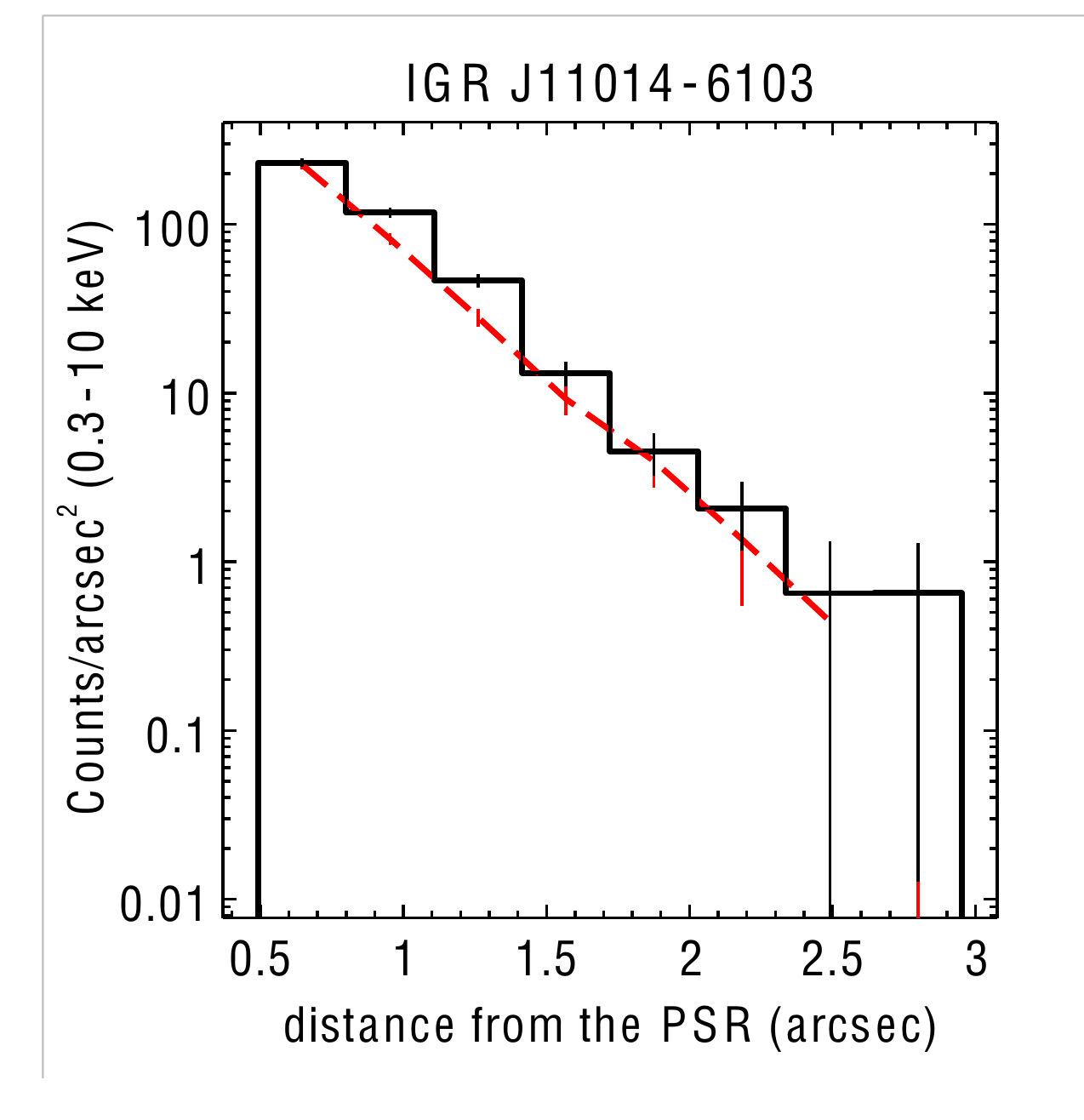}
   \includegraphics[width=0.5\textwidth]{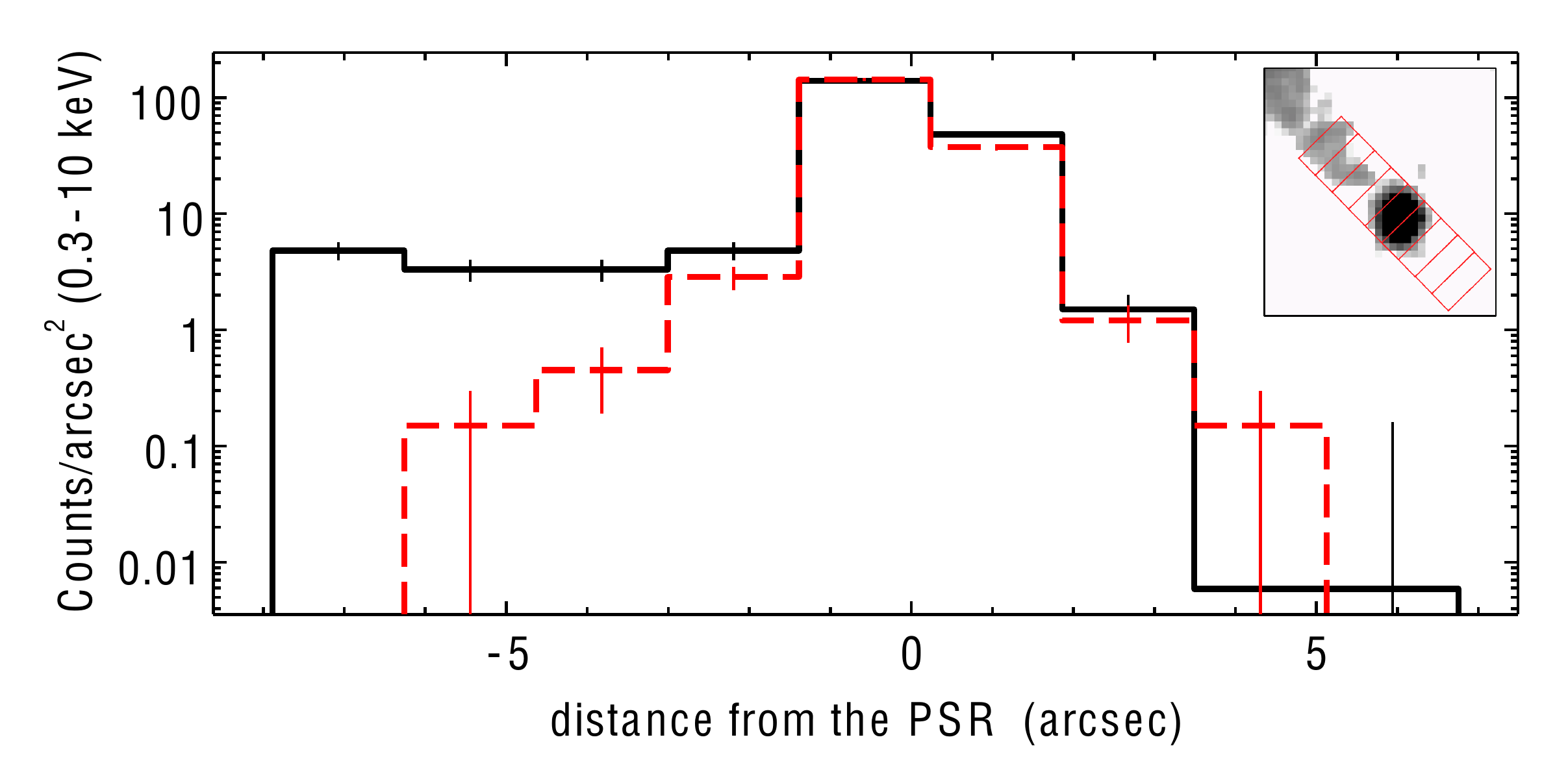}
   \caption{\textit{Top panel:} Radial profile within 3\arcsec\ from
     the source PSR (histogram in black) compared to the PSF (red
     dashed line) simulated with the ChaRT tool at the PSR position
     and with the measured energy spectrum. \textit{Bottom panel:}
     same plot as above, but for the regions marked in the inset
     picture.  Positive radial distances are in front of the source,
     in the SW direction, negative are backwards in the NE direction
     (i.e. in the direction of the PWN).}
   \label{ima:radial_profile}
 \end{figure}

To verify the point-like nature of PSR, we extracted the radial
intensity profile of the source (background subtracted) and compared
it to the corresponding instrumental point-spread function (PSF).  The
PSF was obtained by simulations performed with \texttt{ChaRT}
\citep{chart}, assuming the spectral profile of the PSR source (see
Table~\ref{tab:chanspec}) and reprojected with \texttt{MARX} v.5
\citep{marx} to the observed source position on the ACIS-I3 chip.
The observed and simulated profiles, extracted within a region of
3\arcsec\ around PSR, are displayed in the upper panel of
Fig.~\ref{ima:radial_profile}.  We verified that there are no
variations on the profile between the front and rear regions.
Similarly, we extracted the intensity profile along the axis of the
PWN from a rectangular region centered on the source and extending up
to 7\arcsec\ (Fig.~\ref{ima:radial_profile}, lower panel).  The
comparison with the simulated PSF shows that there is no detectable
extension in front of the source, and that the PWN is smoothly
connected to the point source, emerging 3\arcsec\ away from it
already.

The relatively poor timing resolution of the \chan\ data in timed
exposure mode (3.2~s) did not permit us to analyze the timing
periodicity of the point source.  Previous searches with \xmm\
\citepalias{pavan11} and Parkes \citepalias{tomsick:2012eu} did not
reveal any significant coherent periodicity. We note, however, that in
almost one-third of the known PWN systems there is no clear pulsating
signal from the neutron star \citep[see,
e.g.,][]{Roberts:2004lr}. This can be due to geometrical phenomena,
when the beam of light from the pulsar is oriented in a direction not
intercepting Earth, as well as physical phenomena, if the presence of
a dense region surrounding the pulsar prevent the direct observation
of the neutron star surface and its pulsations
\citep{kargalstev-pavlov-2008}.

\subsubsection{Spectra}
\label{sec:spectra}

\begin{table}
  \caption{Best fit spectral parameters for the \lhn\ components.}
  \center
  \begin{tabular}{@{}ccccc@{}}
    \hline
    \hline
    \noalign{\smallskip} 
    &  $N_{\rm H}$ & $\Gamma$ & $F_{\rm 2-10~keV}$ & $\tilde\chi^2$ /d.o.f.  \\
    &  {\small (10$^{22}$~cm$^{-2}$) }& & {\small (10$^{-13}$~\ferg)} & \\
    \noalign{\smallskip} 
    \hline        
    \noalign{\smallskip} 
    PSR & 1.0$\pm$0.2  & 1.1$\pm$0.2 & 6.1$\pm0.6$  &1.09 / 70 \\
    \noalign{\smallskip} 
    PWN & 0.8$\pm$0.1  & 1.9$\pm$0.1 & 6.7$\pm 0.5$  & 0.82 / 76 \\
    \noalign{\smallskip} 
    Main jet & 0.8$\pm$0.2  & 1.6$\pm$0.2 & 5.4$\pm 0.5$  & 1.08 / 65  \\
    \noalign{\smallskip} 
    \hline
  \end{tabular}
  \tablefoot{All spectra were fit using an absorbed power-law model (photon index $\Gamma$).  
    Uncertainties 
    are at 90\% c.l. on the spectral parameters and 68\% c.l. on the fluxes.}
  \label{tab:chanspec}
\end{table}

We extracted background-corrected spectra of the different components
of the \lhn\ with the standard \texttt{CIAO/specextract} tool, and
analyzed them with \texttt{xspec} \citep[v.12.7.1; ][]{xspec}.\\
All the extended components and PSR were comprised in the FOV of ACIS
I3. The main jet, which was comprised in ACIS I2, was the only
exception.  Therefore, the background regions were chosen from source
free areas in ACIS chips I2 and I3.  We verified that different
reasonable choices of the
background did not significantly affect our spectral results.\\
We extracted the spectrum of PSR from a circular region of 2.3\arcsec\
centered on the source to avoid contamination from the PWN and the
jets. We created the corresponding response matrix file (RMF) and
energy-corrected ancillary response file (ARF) with the \texttt{CIAO}
standard tools.  The spectrum was grouped to have at least 15 counts
per energy bin. For the jet and PWN, we followed standard
\texttt{CIAO} spectral extraction procedure for extended sources. We
used \texttt{specextract} to compute weighted RMFs and ARFs for each
component.

All spectra could be fit with a simple absorbed power-law model, with
best-fit parameters reported in Table~\ref{tab:chanspec}.  We did not
attempt any spectral extraction for the counterjet, given the low
number of counts. In addition to the previous jet analysis, we
extracted spectra from different rectangular regions along the jet, to
search for possible spectral variations. The data, however, did not
reveal any spectral variation between the various regions, within the
uncertainties.
The source PSR showed a power-law photon index of $\Gamma=1.1 \pm
0.2$, which is compatible with that expected from young energetic
pulsars surrounded by their wind nebulae \citep{gotthelf2008}. A
softer spectrum is obtained for the main jet and the PWN, with
$\Gamma=1.6 \pm 0.2$ and $1.9 \pm 0.1$, respectively. For the PWN,
such a relatively soft spectrum is expected (see, e.g.,
\citealp{Gotthelf2004}; and Fig.~8 in \citealp{Li2008}).\\
We also tentatively fitted a pure thermal emission model
(\texttt{vmekal} model in \texttt{xspec}) to the jet spectrum (see,
e.g., the adiabatic expansion scenario for the precessing jets of
SS~433, \citealp{Migliari2002}).
This model provides a good fit to the data, ($\tilde\chi^2 = 1.09 / $
65 d.o.f., similar to that obtained with the power-law model), with
best fit parameters \nh $= (0.8 \pm 0.2) \times 10^{22}$ cm$^{-2}$ and
$kT = 11 ^{+17}_{-5}$~keV.
Although line emission should be expected at such temperatures, in
particular around 6-7 keV, the relatively low count statistics of the
jet spectrum prevents us from probing the existence of such lines.
Our fit results, therefore, cannot clearly distinguish between a pure
thermal and a power-law model (a fit using both components provides a
slightly worse $\tilde\chi^2$ and is unable to constrain neither the
plasma temperature nor the value of \nh). We note, however, that in a
thermal scenario the plasma is expected to cool down on length-scales
much shorter than the $\sim 11$~pc displayed by the main jet, and,
therefore, a continuous reheating of the emitting material would be
required to explain the observed lack of spectral changes along the
jet.
In addition, pulsar jets are supposed to have a magnetohydrodynamical
origin \citep[see, e.g.,][]{Bogovalov1999}, and therefore a power-law
component from synchrotron emission should in any case be expected
\citep[see, e.g.,][]{Pavlov-2003, johnson-wang-10, deluca2011,
  Hui:2012fk}.
We conclude that the thermal model is disfavored with respect to the
power-law model, and a synchrotron scenario is assumed in the
following.

The jet feature and the counterjet together provide almost 1/3 of the
total X-ray flux ($\sim 2 \times 10^{-12}$ \ferg) of the \lhn.

 \begin{figure}[h!]
   \centering
   \includegraphics[height=0.3\textwidth]{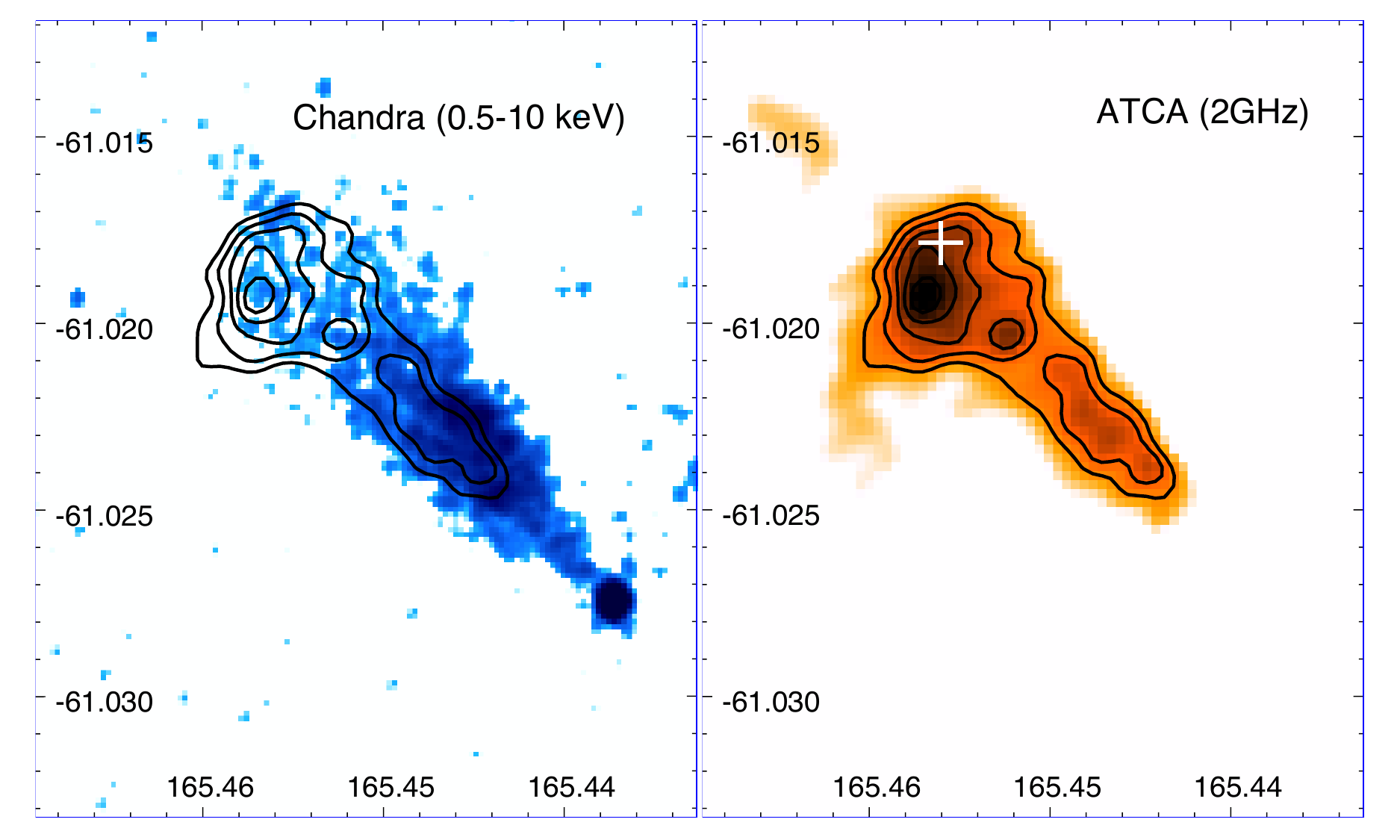}
   \caption{\chan\ (left) and \ATCA\ 2GHz (right) images of the
     PWN. The \ATCA\ contours are overplotted on both images. On the
     right panel, a cross marks the position of the radio source as
     reported in the MGPS-2 survey (see text for details).}
   \label{ima:chandra-atca}
 \end{figure}

\begin{figure}[h!]
  \centering
  \includegraphics[height=0.3\textwidth]{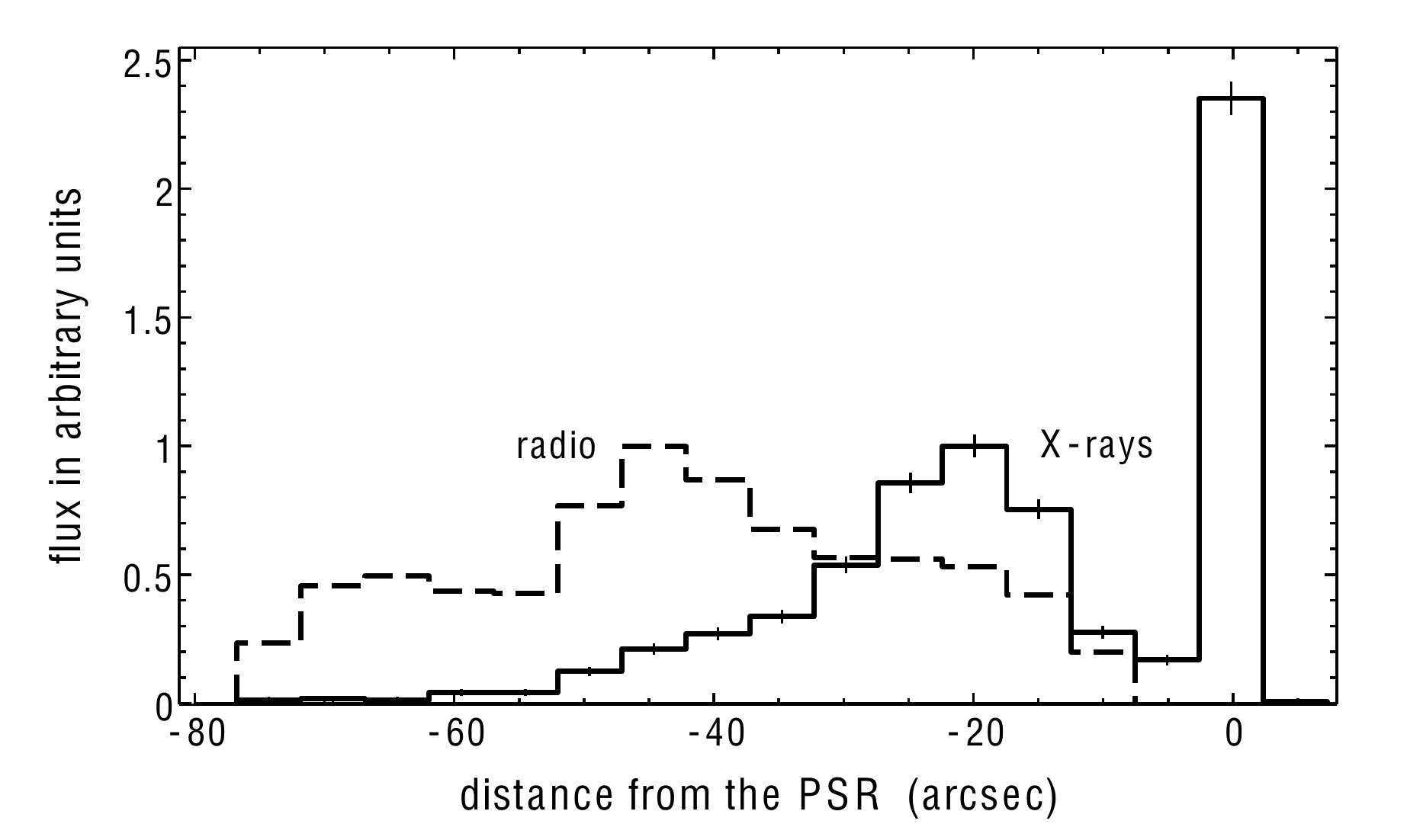}
  \caption{Intensity profiles along the PWN, at 2GHz (dashed line) and
    0.5-10 keV (solid line), as a function of the distance from
    PSR. The radio and X-ray profiles peak at different positions,
    separated by 22\arcsec. This can be explained by the cooling of
    the emitting particles. The shape of the PWN at both wavebands and
    its alignment are in favor of a high pulsar velocity
    ($>$1000~km~s$^{-1}$).}
  \label{ima:IntProfile}
\end{figure}

\subsection{\ATCA\ radio observations}
\label{sec:atca}
We observed the \lhn\ region on January 11, 2013 and February 22, 2013
with \ATCA\ (project C2651), using the new Compact Array Broadband
Backend (CABB) receiver at array configurations of EW352 and 6A, and
wavelengths of 20, 6, and 3~cm ($\nu$=2, 5.5 and 9~GHz), with
bandwidths of 2 GHz. The observations were carried out in
``change-frequency'' mode, totaling $\sim$24 hours of integration over
the two observing periods. The sources \object{PKS~B1934-638} and
\object{PKS~1059-63} were used for primary
and secondary (phase) calibration, respectively.\\
We used the \texttt{miriad} \citep{sault1995} and \texttt{karma}
\citep{karma1996} software packages for reduction and analysis.
Images were formed using \texttt{miriad} multifrequency synthesis
\citep{sault1995} and natural weighting. They were deconvolved using
the \texttt{mfclean} and \texttt{restor} algorithms with primary beam
correction applied using the \texttt{linmos} task. We used a similar
procedure for both U and Q Stokes parameter maps. Because of the good
dynamic range (signal to noise ratio between the source flux and
$3\sigma$ noise level), we applied self-calibration, which resulted in
our best total intensity image (see
Fig.~\ref{ima:chandra-atca}). While our shortest baseline was 46~m (at
EW352 array), we still suffered from the missing flux from the lack of
short (zero) spacings.  This effect was far more pronounced at the
higher frequencies: at $\nu$=5.5 and 9 GHz, the observations were
significantly affected by this ``missing short spacing'' and,
therefore, were excluded from the analysis, which was then performed
only in the 2~GHz band.

The candidate radio counterpart \object{MGPS~J110149-610104}
\citepalias{pavan11}\ was clearly detected in our 2~GHz image as an
extended source (the image resolution is 5.177\arcsec$\times$
4.146\arcsec\ at PA=-8.5828$\degr$ with an estimated r.m.s. noise of
0.15~mJy~beam$^{-1}$). Its extension and position coincide well with
the X-ray contours of the PWN (Fig.~\ref{ima:chandra-atca}), the
source extends up to 80\arcsec\ from PSR (Fig.~\ref{ima:IntProfile}).
Its total integrated flux density is 23$\pm$2~mJy at 2~GHz.
There was no reliable detection in the Q and U intensity parameters at
any observed frequency that could be associated with this object.
Without these detections, the Faraday rotation and consequently the
magnetic field properties could not be determined.

The PSR and the jets were not detected in the 2~GHz image, providing
an upper limit of 0.45~$m$Jy~beam$^{-1}$ at a 3$\sigma$ confidence
level.  For the jet, this translates into a total flux, integrated
over the region used for the X-ray spectral extraction (a boxed region
of size 228.6\arcmin\ $\times$ 40.5\arcmin), lower than $\sim 4 \times
10^{-15}$~\ferg.

\subsection{Simplified precession model}
\label{sec:model}

Our \chan\ observation shows the presence of a jet and a counterjet,
both well collimated, with different brightness and elongation. This
might be interpreted as due to the different orientation with respect
to the line of sight under the assumption that the jet and counterjet
are intrinsically identical in terms of luminosity and extension.  The
data also revealed a clear corkscrew modulation of the main jet,
reminiscent of that seen in other galactic jets \citep[see,
e.g.,][]{Stirling:2002uq,durant2013}, which can be interpreted either
as precession of the neutron star or as kink instabilities in a
ballistic jet (see in Sect.~\ref{sec:discussion}).

\begin{figure}
  \begin{center}
    \includegraphics[width=0.45\textwidth]{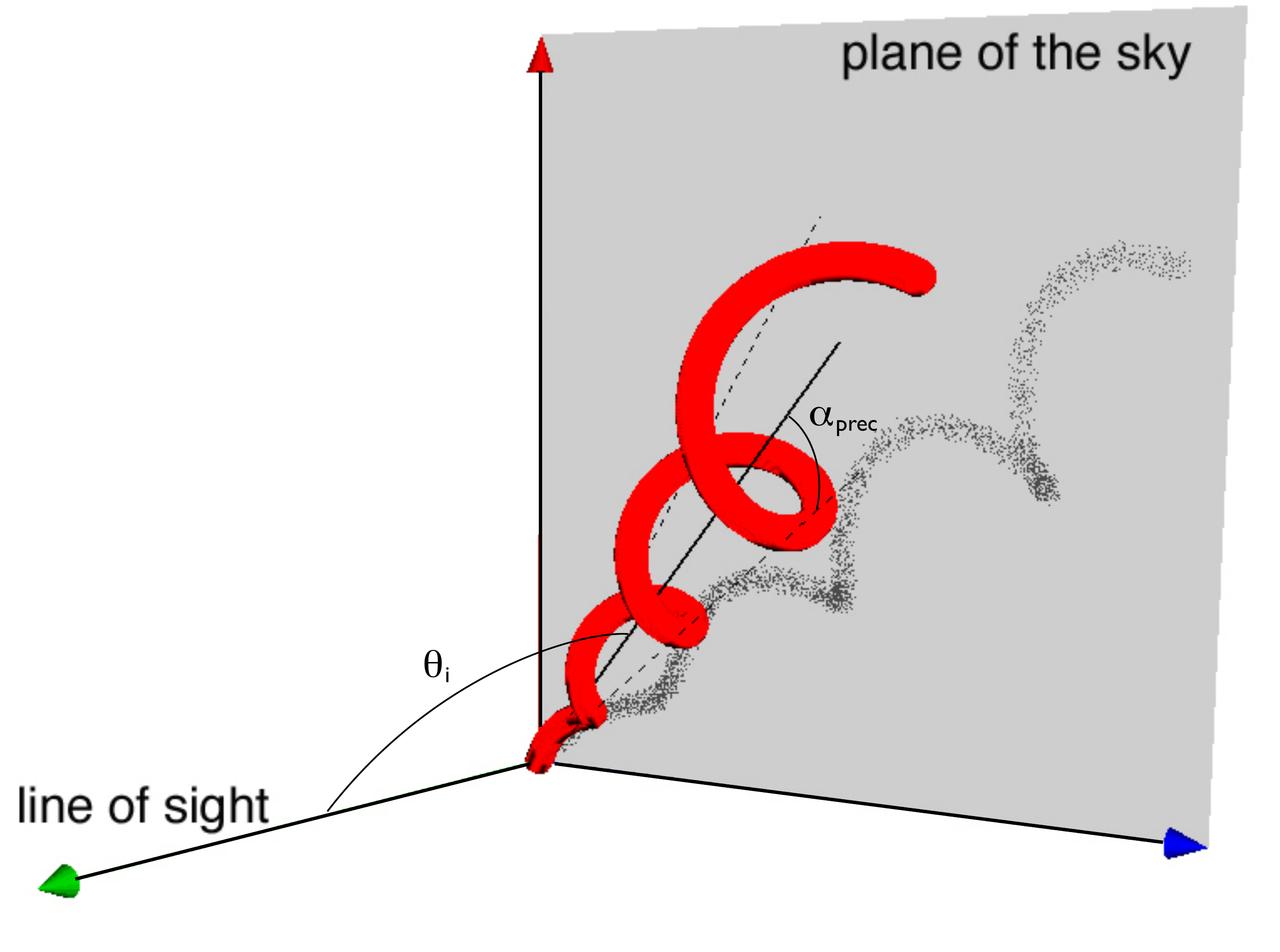}
  \end{center}
  \caption{Sketch of the precession model. The semi-aperture angle of
    the precession cone is marked as $\alpha_{\rm prec}$. The angle
    between the jet axis and the line of sight $\theta_{\rm i}$ is
    also shown. The projection of the 3D helix on the plane of the sky
    (shown here in grey) is used to fit the \chan\ image after being
    corrected for the Doppler factors of each particle.}
  \label{ima:model}
\end{figure}

To further investigate this scenario, we used a simplified precession
model, which describes the large-scale structure of the jet.  We
considered particles moving linearly away from the central position,
each of them launched toward a direction lying on the precession cone
surface (see Fig.~\ref{ima:model}).  In the absence of any external
effect, the interpolation of all particles forms a conical helix.  We
let the semi-aperture angle of the cone, the period of precession, the
phase angle of the helix, and the velocity of the particles moving
along the helix, as free parameters. In this model, all particles move
at the same velocity, equal to the bulk velocity of the jet.  The
section of the helix has a 2D gaussian profile to simulate the degree
of collimation of the jet. The width of this gaussian is another free
parameter.\\
All of the model parameters are strongly dependent on the particles'
velocity $\beta\, c$ (where $c$ is the speed of light), therefore, we
kept this parameter frozen and performed simulations for different
values of $\beta$.  The velocity vector of each particle is used to
determine the Doppler boosting factor of the emitted radiation $\delta
= \gamma^{-1} \, (1-\beta \cos\theta_{\rm i})^{-1}$ (where $\gamma =
(1-\beta^2)^{-1/2} $ is the Lorentz factor of the particle and
$\theta_{\rm i}$ is the angle to the observer).

\begin{figure}[h!]
   \centering
   \includegraphics[width=0.5\textwidth]{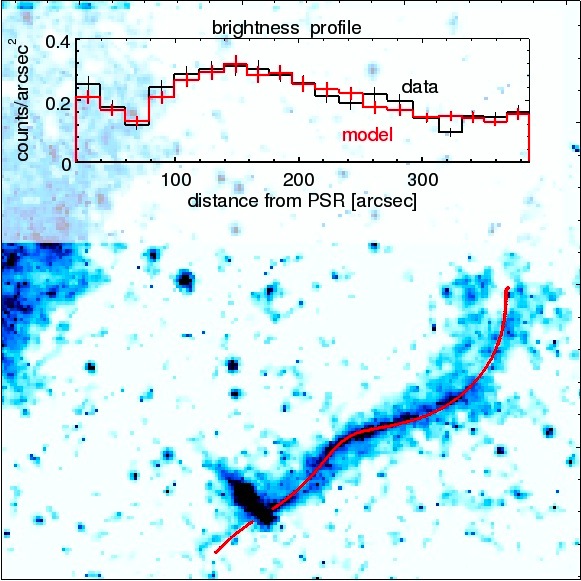}
   \caption{Best-fit model of a precessing jet (solid red line)
     overplotted on top of the \chan\ image of the \lhn.  The plot
     (and the inset) is relative to particles with bulk velocity
     $\beta =0.8$ (see text for details).  The inset shows the
     brightness profile measured along the jet (in black), and the
     brightness profile obtained from the model (in red in the online
     version).}
   \label{ima:precessionModel}
 \end{figure}

We fixed the position of the jet base in image coordinates to match
the observed launching point of the jet.
The 3D direction of the precession axis is constrained by the ratio
$L_{counterjet}/L_{jet} \sim 0.05$ observed between jet and counterjet
apparent luminosities. In this scenario, the $L_{counterjet}/L_{jet}$
ratio is due to the different orientation of the two jets, and to the
relativistic beaming effects due to the bulk motion of the emitting
particles.
Following the same Doppler boosting prescription as above, at each
bulk velocity $\beta$ we find the corresponding angle between the jet
axis and the line of sight $\theta_{\rm i}$ needed to reproduce the
observed ratio $L_{counterjet}/L_{jet}$.  We then derive a lower limit
for the velocity of the particles along the jet axis $v_{\rm
  jet, \parallel} \gtrsim 0.52~c$~(sin\,$\theta_{\rm i}$)$^{-1}$.
The projection of the helix on the plane of the sky, corrected for the
Doppler boosting factor of each particle, is then fit to the data with
the Sherpa package \citep{sherpa}, using the Nelder-Mead Simplex
optimization method with Cash statistics (see
Fig.~\ref{ima:precessionModel}).  We explored several values of the
bulk velocity in the permitted range $\beta = v_{\rm jet, \parallel}/c \, \ge 0.52\,$. \\
After the 2D spatial fit is performed, we compare the brightness
modulation observed in our \chan\ data along the jet (see inset in
Fig.~\ref{ima:precessionModel}) to the modulation obtained from the
model, which is given by approaching/receding relativistic beaming
effects, and therefore, provides three-dimensional information of the
jet morphology.

The minimum bulk velocity $\beta = 0.52$, which corresponds to a
face-on jet with $\theta_{\rm i}$=0, is excluded by the observed
elongation of the jet and the presence of the counterjet in the \chan\
data. The fit also provides a more stringent limit $\beta \gtrsim
0.7$, since an inclination angle $\theta_{\rm i} \gtrsim 40 \degr$ is
required to match the jet morphology and the Doppler boosting effects
on the brightness profile.  Within this refined range, a value $\beta
\sim 0.8$ is favored (see Fig.\ref{ima:precessionModel}), resulting in
a jet inclined by $\theta_{\textrm i} \sim 50 \degr$ to the line of
sight (with an intrinsic length of the main jet of $\sim$15 pc), a
semi-aperture angle of the precessing cone of 4.5$\degr$ and a
precession period of nearly 66~yr. The relatively low signal to noise
ratio of the image did not permit us to estimate the corresponding
uncertainties.

 \section{Discussion}
 \label{sec:discussion}
Besides the pulsar, the PWN, and the perpendicular jet-like feature
 described in earlier works, our \chan\ observation has revealed the
 presence of a counterjet extending in the opposite direction from
 that of the main jet for about 1.5\arcmin. The \chan\ data also
 showed for the first time that the long jet is highly collimated and
 displays a strikingly well-defined and continuous corkscrew
 modulation, with a length of $\sim$5.5\arcmin.  The column density
 estimated in X-rays for the different components of the \lhn\
 ($N_{\rm H} = [0.8 \pm 0.2] \times 10^{22}$~cm$^{-2}$, see
 Table~\ref{tab:chanspec}) is in good agreement with that found in
 various regions of the neighboring \SNR\ \citep[ {$[0.4-0.7] \times
   10^{22}$~cm$^{-2}$};][]{Garcia:2012fk}, proving the distance of the
 \lhn\ to be \mbox{$\sim 7$~kpc $(d_7)$.}

 The radio source \object{MGPS~J110149-610104}, previously identified
 as a possible counterpart of the \lhn\ \citepalias{pavan11}, is
 clearly detected in our \ATCA\ observation.  Its morphology closely
 matches the X-ray PWN (see Fig.~\ref{ima:chandra-atca}). The flux
 density measured at 2~GHz ($23 \pm 2$ mJy) is compatible with the
 flux density reported at 843~MHz in the MGPS-2 archive
 \citepalias[$24 \pm 5$ mJy; see][]{pavan11}. This indicates that the
 tail has a flat radio spectrum, i.e., with radio spectral index
 $\alpha \sim$ 0, as typically observed in PWNe \citep{gaensler2006}.
 The neutron star and jet structures are not detected in radio (upper
 limit at $45$~mJy~beam$^{-1}$), which is a common feature amongst
 many high-energy emitting pulsars \citep{fermiPSRcat2010} and pulsar
 jets \citep{Dodson2003,Bietenholz2004}.
 The source PSR in \j11\ shows standard properties of rotation powered
 pulsars and their associated PWNe \citep{Possenti2002,Gotthelf2004,
   Li2008}, in terms of X-ray luminosity, power-law photon index, and
 stability over at least 30~yr \citepalias{pavan11}, even if
 pulsations have not been detected so far \citepalias[see
 Sect.~\ref{sec:psr};][]{pavan11,tomsick:2012eu}.
 The similar luminosities inferred for the PSR and PWN are also in
 agreement with several observed PWN systems, and with the estimated
 system age of $1-2 \times 10^{4}$~yr \citep[][]{Li2008}.
 Furthermore, the pulsar spin-down power $\dot{E}_{\rm PSR}$ can be
 inferred from the total observed X-ray luminosity of the system
 \citep[see, e.g.,][]{Li2008}, $L_{\rm X}\approx 1.2 \times 10^{34}
 \,\, d_7^{2}$~erg~s$^{-1}$.  This results in a rather high spin-down
 power $\dot{E}_{\rm PSR} \sim 10^{37}$~erg~s$^{-1}$ for the PSR in
 \j11, in agreement with the observation of a bright PWN
 \citep{gaensler2006, Li2008}.  For a system with this spin-down
 power, the presence of the parent SNR should also be expected
 \citep{gaensler2006}.

 The PWN in the \lhn\ displays a clear symmetry axis that defines the
 direction of motion of the system, pointing toward the center of the
 nearby \SNR\ (see Fig.~\ref{ima:flat-fielded}). This alignment,
 together with the similar distance derived from our $N_{\rm H}$
 measure, further support the proposed association between the two
 objects \citepalias{tomsick:2012eu}, and imply a pulsar
 speed\footnote{The velocity derived here for \j11\ differs from the
   one reported in \citetalias{tomsick:2012eu}, as we adopted the
   refined distance of 7~kpc for \SNR\ reported in
   \citet{Garcia:2012fk}} $v_{\rm PSR} \sim (1100 - 2200)\
 d_7$~km~s$^{-1}$. This kick velocity makes \j11\ one of the most
 extreme runaway pulsars known so far.
 Further support for this high velocity comes from the observed
 geometry of the PWN.  Following the analytical description of
 bow-shock wind confinements proposed by \citet{wilkin2000}, a
 velocity of $\gtrsim 1000$~km~s$^{-1}$ is required to match the PWN
 morphology, where the following parameters were adopted:
 $\dot{E}_{\rm PSR} \sim 10^{37}$~erg~s$^{-1}$, an external medium
 particle density of $n_{\rm ISM} = 0.1$~cm$^{-3}$, and an inclination
 $i = 0 \degr$ of the velocity vector of the pulsar \emph{(not the
   inclination of the jet axis)} with respect to the plane of the sky.

 The observed shift between the X-ray and radio maxima ($\theta_{\rm
   peaks} \sim 22\arcsec$, see Fig.~\ref{ima:chandra-atca}) can be
 used to estimate the PWN magnetic field. Taking a minimum PWN
 backflow velocity equal to the pulsar's speed $v_{\rm PSR} \approx
 1000~$km~s$^{-1}$, it takes a time $t_{\rm peaks} \sim \theta_{\rm
   peaks}\,d / v_{\rm PSR}$ for particles to travel from the X-ray
 peak to the radio peak ($d$ is the distance to \j11).  If the same
 particle population is responsible for the emission at both energies,
 the synchrotron loss timescale is constrained to be $t_{\rm peaks}
 \gtrsim t_{\rm sync}(E_{\rm e^{-}}, B_{\rm PWN}) \approx 422 \,
 E_{\rm e^{-}}^{-1} \, B_{\rm PWN}^{-2} $~s, where $E_{\rm e^{-}}$ is
 the energy of the emitting electrons and $B_{\rm PWN}$ is the
 magnetic field in the nebula.  Synchrotron emission at frequencies
 $h\nu_{\rm c} = 5$~keV requires, on the other hand, that $h\nu_{\rm
   5keV} \approx 5.2 \,B_{\rm PWN}\,E_{\rm e^{-}}^{2}$. Putting these
 constraints together, a minimum value $B_{\rm PWN} \approx 10-20\
 \mu$G is required, in good agreement with known PWN magnetic fields
 \citep{gaensler2006}.
  
 The PWN X-ray and radio spectra obtained in
 Sect.~\ref{sec:observations} imply the existence of a break frequency
 in between the two energy bands.  Extrapolating the radio and X-ray
 power-law-like spectra, such a break should fall at $\nu_{\rm br
 }\sim 5 \times 10^{11}$~Hz. This break frequency could be used, in
 principle, to further constrain the emitting electron distribution
 and/or the PWN magnetic field. However, the spectral break $\Delta
 \alpha \sim 0.92$, is much larger than the expected value of $ 0.5$
 in a single-population synchrotron-emission scenario. A similar
 situation is observed in several PWN systems in which additional
 (still unclear) processes may need to be considered for an accurate
 derivation of the properties of the electron distribution and the
 nebular magnetic field (\citealp{gaensler2006}; see also
 \citealp{Reynolds2009}).

 The distance $d \approx 7$ kpc derived for the \lhn\ implies an
 intrinsic length of the main jet $l_{\rm jet} \, \ge \,11$~pc, making
 it the longest X-ray jet detected so far in our Galaxy (accounting
 for projection further increases this estimate by a factor
 ($\sin\,\theta$)$^{-1}$ , where $\theta \in [0,\pi/2]$ is the jet
 angle to the line of sight, with a most likely value $\theta \sim 50
 \degr$, see Sect.~\ref{sec:model}). The observed precession-like
 helical modulation, similar to that seen, e.g., in other galactic
 jets \citep{Stirling:2002uq,durant2013}, promptly suggests a
 ballistic jet origin for this structure.
 The different X-ray brightness and elongation of the jet and
 counterjet can also be easily reconciled in this scenario, owing to
 their different orientation with respect to the observer.  To further
 characterize the ballistic pulsar jet scenario, we used a simplified
 helical model describing the large-scale structure of the jet.  The
 model describes the data well, both in terms of the spatial
 modulation and the jet brightness profile (see
 Fig.~\ref{ima:precessionModel}), thus confirming the
 three-dimensional helical structure and consequently the truly
 ballistic jet nature of the feature.  An interpretation in a
 diffusion scenario of particles in the interstellar medium
 \citep{bandiera2008} can be excluded, as it would require an
 underlying helical structure for the interstellar medium's magnetic
 field, which is instead known to be dominated by its turbulent
 component at scales below $\sim 100$~pc \citep{gaensler2011,
   Giacinti:2012fk}.

 The helical pattern seen along the jet could be explained either by
 free precession of the pulsar (with a period of $\sim 66$~yr; see
 Sect.~\ref{sec:model}) or by the development of kink instabilities
 \citep[see, e.g.,][]{Lyubarskii1999}.  In this latter case, the
 period observed in the helical model would be associated with the
 timescale at which successive kinks appear. However, a low level of
 periodicity should be expected in this scenario, unless the
 instability is triggered by precession \citep{durant2013}. Since our
 data reveal only approximately 1.5 helical steps along the main jet,
 such periodicity level in the trigger of the kink instabilities is
 difficult to assess.  Periodic modulation has, on the other hand,
 been tentatively interpreted in some systems as free precession of
 the pulsar \citep[see, e.g.,][and references therein]{akgun2006,
   jones2012,durant2013}, with periods up to several years (see,
 e.g. \citealp{weisberg2010}).  We note, however, that periods as long
 as several tens of years could hardly be detected in systems which do
 not show an extended jet, as in these systems the periodic modulation
 must be inferred by variations of the pulsar properties (e.g.,
 spectral and timing properties, or of the linear polarization angle),
 based on monitoring
 of the source typically over timescales of a few years.\\
 A free pulsar precession could be explained assuming a given value
 for the oblateness ($\epsilon$) of the rotating neutron star
 \citep[see e.g.][]{jones2012}, which can be estimated from the
 pulsar's spin and the precession period as $\epsilon = (I_3 -
 I_1)/I_1 = P_{spin}/P_{prec}$ \citep[where $I_1$ and $I_3$ are
 principle moments of inertia; see, e.g.,][]{haberl2006}.  For
 comparison with the values inferred for several likely precessing
 isolated pulsars, in the range $\epsilon \approx$ $10^{-4}$ to
 $10^{-10}$ \citep{Jones2001,jones2012,durant2013}, we compute here
 the oblateness of \j11.  A rough estimate for the spin period of PSR
 is obtained considering the pulsar age (10-20 kyr) and spin-down
 energy estimated above, $\dot{E}_{\rm PSR} \sim
 10^{37}$~erg~s$^{-1}$, which provides $P_{spin}$ of the order of
 0.1~s.  A low value of the oblateness parameter is, therefore,
 obtained for the pulsar in the \lhn, $\epsilon \sim 0.5 \times
 10^{-10}$, which is close to values found in other cases (see above).

 In our simulations, we did not consider any bending of the precessing
 axis, since it is not apparent in our \chan\ images. The lack of
 bending implies that the curvature radius of the jet has to be large,
 $R_{\rm curv} \gtrsim l_{\rm jet}$. Since $R_{\rm curv} \propto
 \dot{E}_{\rm jet} \, R_{\rm jet}^{-2} \, \rho_{\rm ISM}^{-1} \, v_{\rm
   PSR}^{-2}$, where $\dot{E}_{\rm jet}$ is the jet power, $R_{\rm
   jet}$ its radius, and $\rho_{\rm ISM}$ is the medium mass density
 (see, e.g., \citealp{Soker2006} and references therein), it follows
 that the jet has to be rather powerful and/or collimated.
 A high jet power is indeed supported by the relative distribution of
 the total X-ray luminosity of the system. The jet contribution is
 similar to that of the pulsar and the PWN, suggesting that a sizable
 fraction of $\dot{E}_{\rm PSR}$ is channeled through the jets.\\
 The jet X-ray spectrum suggests a synchrotron origin for its emission
 (see Sect.~\ref{sec:observations}), with a relatively hard photon
 index \mbox{$\Gamma = 1.6\, $.}  Assuming that the radio and X-ray
 emission are produced by a single electron population, a straight
 extension of the X-ray spectrum would yield a flux $\sim
 10^{-16}$~erg~cm$^{-2}$~s$^{-1}$ at 2~GHz, for a bandwidth of 2
 GHz. This estimate, being lower than the upper limit $\sim 4 \times
 10^{-15}$~\ferg\ obtained with ATCA at the same frequency, can
 naturally explain the radio nondetection of the jet.

 The absence of breaks in the X-ray spectrum all along the jet implies
 that no significant cooling of the X-ray emitting electrons is taking
 place. We use the condition $ h\nu_{\rm sync}/5$keV $\equiv h\nu_{\rm
   5keV} \approx 5.2 \,B_{\rm jet}\,E_{\rm e^{-}}^{2}\, $ again,
 require that the synchrotron timescale $t_{\rm sync}$ is at least
 equal to the time taken by electrons to travel up to the jet tip,
 $\sim l_{\rm jet}/v_{\rm jet}$, and thereby obtain a maximum value
 for the jet magnetic field $B_{\rm jet}\lesssim\,65 \, (h \nu_{\rm
   5keV})^{-1/3} \, (l_{\rm jet}/$15~pc$)^{-2/3}\,(v_{\rm
   jet}/0.8c)^{2/3}$~$\mu$G.  A further constraint on $B_{\rm jet} $
 is derived from the requirement that electrons are confined within
 the jet so that their Larmor radius $r_{\rm L} \approx 1.75 \times
 10^{25} \, E_{\rm e^{-}} \, B_{\rm jet}^{-1} $~cm is at most equal to
 $R_{\rm jet}\sim 2.3 \times 10^{17}\,d_{7}$~cm.  Putting these
 constraints together, the jet magnetic field is found to be $B_{\rm
   jet} \sim $ 10 -- 65~$\mu$G.  Assuming instead near-equipartition
 between particles and magnetic fields (see, e.g.,
 \citealp{Pacholczyk1970}), we find $B_{\rm jet}^{\rm equip} \approx
 15$~$\mu$G, implying a minimum jet power $\dot{E}_{\rm jet}^{\rm
   equip} \approx 2 \times 10^{35}$~erg~s$^{-1}$. This estimate does
 not account, however, for the presence of relativistic protons, if
 any,c and/or thermal material in the jet, which would make the total
 jet kinetic luminosity higher, in line with the high relative
 contribution (nearly 1/3) of the jets to the total system's
 luminosity and with the absence of any noticeable jet bending.

 Although jet-launching mechanisms in pulsars are not yet fully
 understood, they likely have a magneto-hydrodynamical origin
 \citep{komissarov2004}.  The ultrarelativistic speed and the
 anisotropy of the pulsar wind \citep{Bogovalov2002} suggest that the
 jets are formed by magnetic hoop stresses onto the wind material
 downstream of the termination shock \citep{Lyubarsky2002}. This
 termination shock, for the same conditions of the external medium, is
 located much closer to the neutron star in a high-velocity pulsar as
 compared to slow-moving systems \citep{gaensler2006}.
This could explain the relatively high jet power and collimation
 degree in the \lhn, compared, e.g., to those from the Crab
 \citep{Weisskopf-2000} and Vela pulsars \citep{Pavlov-2003}.

 \section{Concluding remarks}
 \label{sec:conclusions}
 The \lhn\ is the first case in which a jet from a hyperfast runaway
 pulsar can be clearly identified.  The associated \SNR\ is the
 remnant of a core-collapse explosion from a massive star
 \citep{Garcia:2012fk}. These types of supernovae are expected to
 produce in some cases neutron stars with natal kick velocities up to
 or exceeding 1000 km~s$^{-1}$ \citep{wang-lai2007,Janka-2012}.
 Simulations of core-collapse mechanisms however have difficulties in
 predicting orthogonal alignment between the pulsar spin-axis and the
 direction of motion for the highest-velocity systems, as observed in
 \j11\ \citep{wang-lai2007}.  Alternative models accounting for this
 misalignment have been suggested \citep[see,
 e.g.,][]{Colpi2002,davies2002}, 
 but these require an extreme rotation of
 the iron core of the presupernova star, not commonly accepted to be
 achievable \citep{Fryer-2004}.

 We remark that for the Guitar nebula \citep{Hui:2012fk}, the pulsar's
 jet nature of the X-ray linear feature is not proven yet, and the
 originating SNR is not known so far. Nevertheless, if the feature
 will be confirmed as a true pulsar's jet, the supernova responsible
 for the system could already constitute the second example for such
 an unusual core-collapse episode, similar to the one originating the
 \lhn.

 Core-collapse supernovae are also expected to produce bipolar
 outflows during or right after the explosion
 \citep{davies2002,paragi2010,Janka-2012}.  These events can, under
 certain circumstances, even lead to the formation of long-duration
 gamma-ray bursts \citep{davies2002,paragi2010, Soderberg2010}.  The
 \SNR\ shows a clear bipolar asymmetry along the NW-SE direction
 \citep{Garcia:2012fk}, which could be the imprint of the past outflow
 activity during the supernova explosion
 \citep{paragi2010,Janka-2012}. The high kick velocity of \j11\ and
 the rough alignment of its jets with the direction of the bipolar
 asymmetries in \SNR\ could then be the echoes of a quenched
 long-duration gamma-ray burst.

 \begin{acknowledgements} This paper is based on \chan\ and \ATCA\
   observations.  We thanks prof.~D.~Lai for useful discussion, and
   the Chandra Uplink Support team, in particular Dr.~J.~Connelly, for
   the support in preparing the observation.  L.P. thanks the
   Soci\'et\'e Acad\'emique de Gen\`eve and the Swiss Society for
   Astrophysics and Astronomy for travel grants sustaining the ongoing
   collaboration between the ISDC and the ATCA team.
 \end{acknowledgements}

\end{document}